\documentclass[11pt]{article}
\usepackage{graphicx,amsmath,amsthm,mathrsfs,amssymb}
\usepackage[usenames]{color}
\usepackage{ulem}
\usepackage{authblk}
\usepackage{cite}
\usepackage{appendix}

\newcommand{\ee}{\end{equation}}
\newcommand{\word}[1]{\,\,\mbox{#1}\,\,}
\newcommand{\reff}[1]{(\ref{#1})}
\newcommand{\beq}{\begin{equation}}
\newcommand{\eeq}[1]{\label{#1}\end{equation}}
\newcommand{\beqa}{\begin{eqnarray}}
\newcommand{\eea}{\end{eqnarray}}
\newcommand{\eeqa}[1]{\label{#1}\end{eqnarray}}
\newcommand{\beg}{\begin{equation*}}
\newcommand{\eeg}{\end{equation*}}

\newcommand{\m}{\!-\!}

\newcommand{\bsplit}{\begin{split}}
\newcommand{\esplit}{\end{split}}

\title{Critical gravity from four dimensional scale invariant gravity}

\author[1]{Ariel Edery\thanks{aedery@ubishops.ca}}
\author[2]{Yu Nakayama\thanks{yu.nakayama@rikkyo.ac.jp}}

\affil[1]{Department of Physics, Bishop's University, \\2600 College Street, Sherbrooke, Qu\'{e}bec, Canada, J1M 1Z7 .\vspace{8mm}}
\affil[2]{Department of Physics, Rikkyo University, Toshima, Tokyo 171-8501, Japan}

\begin{document}
\date{}
\maketitle
\begin{abstract}
We show that a critical condition exists in four dimensional scale invariant gravity given by the pure quadratic action $\beta \,C_{\mu\nu\sigma\rho} C^{\mu\nu\sigma \rho} + \alpha \,R^2$ where $C^{\mu}_{\,\,\nu \sigma \rho}$ is the Weyl tensor, $R$ is the Ricci scalar and $\beta$ and $\alpha$ are dimensionless parameters. The critical condition in a dS or AdS background is $\beta =6 \alpha$. This leads to critical gravity where the massive spin two physical ghost becomes a massless spin two graviton. In contrast to the original work on critical gravity, no Einstein gravity with a cosmological constant is added explicitly to the higher-derivative action. The critical condition is obtained in two independent ways. In the first case, we show the equivalence between the initial action and an action containing Einstein gravity, a cosmological constant, a massless scalar field plus Weyl squared gravity. The scale invariance is spontaneously broken. The linearized Einstein-Weyl equations about a dS or AdS background yield the critical condition $\beta=6\alpha$. In the second case, we work directly with the original quadratic action. After a suitable field redefinition, where the metric perturbation is traceless and transverse, we obtain linearized equations about a dS or AdS background that yield the critical condition $\beta= 6\alpha$. As in the first case, we also obtain a propagating massless scalar field. Substituting $\beta=6\alpha$ into the energy and entropy formula for the Schwarzschild and Kerr AdS or dS black hole in higher-derivative gravity yields zero, the same value obtained in the original work on critical gravity. We discuss the role of boundary conditions in relaxing the $\beta=6\alpha$ condition.
\end{abstract}

\setcounter{page}{1}
\section{Introduction}
Critical gravity was introduced in \cite{PopeLu1} where the following four-dimensional action was considered: 
\beq
S=\dfrac{1}{2\kappa^2}\int \sqrt{-g}\,d^4x \,(R-2\Lambda + b_1\, R_{\mu\nu}R^{\mu\nu} + b_2\,R^2) 
\eeq{S1} 
where $R$ is the Ricci scalar, $\Lambda$ the cosmological constant, $R_{\mu\nu}$ the Ricci tensor and $b_1$, $b_2$ and $\kappa$ are constants. This theory describes a massless spin two graviton, a massive spin two ghost and a massive scalar \cite{Stelle1, Stelle2}. The massive scalar mode was eliminated via the choice $b_1=-3\,b_2$ leading to conformal gravity for the higher-derivative part. Most importantly, it was found that if the parameter $b_2$ was chosen to be $-\frac{1}{2 \Lambda}$ then the massive spin two ghost becomes a massless spin two graviton. This was dubbed critical gravity. Besides the massless spin two graviton, critical gravity contains logarithmic spin two modes which are ghosts \cite{Porrati}. Around the same time, it was found that the massive spin two ghost of pure Weyl squared gravity can be eliminated via a suitable boundary condition \cite{Maldacena}. The role of boundary conditions in extending critical gravity as well as eliminating the logarithmic spin two modes were then discussed in \cite{PopeLu2}. We will return to the role of boundary conditions later. For now, we focus on the critical condition. 

In this paper, we show that one can obtain a critical condition with only the higher-derivative part of the action where Einstein gravity with a cosmological constant is not included explicitly. This insight stems from recent work on pure $R^2$ gravity that shows that it is equivalent to Einstein gravity with a cosmological constant plus a massless scalar field \cite{Lust1,Lust2,YNAE1,Lust3,YNAE2} (the massless scalar is absent in the Palatini formalism \cite{YNAE2A}. See also other work on $R^2$ gravity in \cite{Rinaldi1,Rinaldi2}). Therefore, one can consider solely the scale-invariant four-dimensional quadratic action 
\beq       
S=\int \sqrt{-g} \,d^4x \,(\beta \,C_{\mu\nu\sigma\rho} C^{\mu\nu\sigma \rho} + \alpha \,R^2)
\eeq{S2}
where $C^{\mu}_{\,\,\nu \sigma \rho}$ is the Weyl tensor, $R$ is the Ricci scalar and $\beta$ and $\alpha$ are dimensionless parameters. We show that this theory has a critical condition at $\beta=6\alpha$. In other words, when this condition is satisfied, the massive spin two ghost becomes a massless spin two graviton. We show this using two independent approaches. In the first approach, we convert the $R^2$ part of the action into its equivalent form involving Einstein gravity with a cosmological constant plus a massless scalar field. This requires a conformal transformation which does not affect the Weyl squared part of the action. One therefore ends up with an Einstein-Weyl action that includes a cosmological constant $\Lambda$ and a massless scalar. The scale invariance is spontaneously broken \cite{YNAE2}. We linearize the Einstein-Weyl equations of motion about a de Sitter (dS) or anti-de Sitter (AdS) background. In our parameterization of the action \eqref{S2}, we find that the critical condition is $\beta=6\alpha$.
In the second approach, we work directly with the original pure quadratic action \reff{S2} and linearize the equations of motion in a dS or AdS background. Upon a field redefinition of the metric perturbation where it is traceless and transverse, the linearized equations lead directly to the critical condition $\beta=6\alpha$. We also obtain a massless scalar from the trace part of the metric perturbation. The two approaches therefore agree in yielding the same critical condition and having a propagating massless scalar. When we substitute the condition $\beta=6\alpha$ into the energy and entropy formula for a Schwarzschild or Kerr AdS or dS black hole in higher-derivative gravity, we obtain zero in agreement with the original work on critical gravity \cite{PopeLu1}. 

Critical gravity suffers from two issues. First, the theory is empty in the sense that it leads to zero energy. Secondly, as already mentioned, the logarithmic spin two modes are ghosts. Motivated by the work of Maldacena \cite{Maldacena}, both of these problems were resolved in \cite{PopeLu2} via boundary conditions. In particular, in \cite{PopeLu2} they were able to extend critical gravity by relaxing the critical condition so that one obtains positive energy solutions. Similarly, we obtain positive energy solutions by imposing boundary conditions and relaxing the $\beta=6 \alpha$ condition.   

\section{Spontaneous symmetry breaking of four dimensional scale-invariant gravity}
Scale-invariant gravity can be expressed by the general four dimensional higher-derivative quadratic action: 
\begin{align}
S_1 = \int d^4x \sqrt{-g} \left(\beta \,C_{\mu\nu\sigma\tau}C^{\mu\nu\sigma\tau} + \alpha R^2 \right) \,.
 \label{Rw1}
\end{align}
The other two terms, Euler density (Gauss-Bonnet term) and the Pontryagin density, are topological in four dimensions and can be ignored in the classical equations of motion. Thus \eqref{Rw1} is the most general classical scale invariant action constructed out of the metric.

The above action is not only scale-invariant but invariant under a larger symmetry; it is restricted Weyl invariant \cite{YNAE3,YNAE1,YNAE2} i.e. it is invariant under the transformation 
\beq
g_{\mu\nu}\rightarrow \Omega^2 g_{\mu\nu}\quad,\quad \word{ with} \Box \Omega=0
\eeq{Rw2}
where the conformal factor $\Omega(x)$ is a real smooth function. 
This symmetry forbids an Einstein-Hilbert term as well as a cosmological constant in \reff{Rw1}. However, we will see how they appear after the symmetry is spontaneously broken. Introducing the auxiliary field $\varphi$, we can rewrite the above action into the equivalent form 
\begin{align}
S_2 =\int d^4x \sqrt{-g} \Big[\beta \,C_{\mu\nu\sigma\tau}C^{\mu\nu\sigma\tau}-\alpha(c_1 \varphi + R )^2 + \alpha R^2 \Big]  
\label{Sb}
\end{align} 
where $c_1$ is an arbitrary constant\footnote{In the path integral formulation, the squared term yields a Gaussian integral over $\varphi$ and does not affect anything i.e. $\int\mathcal{D}\varphi e^{- i\alpha\,c_1^2 \int d^4x\sqrt{-g}(\varphi - f(x))^2} = \mathrm{const}$.}. Expanding the above action we obtain  
\begin{align}
S_3 &= \int d^4x \sqrt{-g} \Big(\beta \,C_{\mu\nu\sigma\tau}C^{\mu\nu\sigma\tau} -c_1^2\,\alpha \,\varphi^2 - 2\alpha c_1 \varphi R \Big) \ . 
\label{Sc}
\end{align}
Action \reff{Sc} is equivalent to the original action \reff{Rw1} and is restricted Weyl invariant as long as $\varphi$ transforms accordingly; it is invariant under the transformations $g_{\mu\nu}\rightarrow \Omega^2 g_{\mu\nu}$, $\varphi\rightarrow \frac{\varphi}{\Omega^2}$ with $\Box \Omega=0$. 

After performing the conformal (Weyl) transformation
\beq
g_{\mu\nu} \to \varphi^{-1}g_{\mu\nu}
\eeq{Conf}
the above action reduces to an action that contains an Einstein-Hilbert term with a cosmological constant:   
\begin{align}
S_4&=\int d^4x \sqrt{-g} \,\Big(\beta \,C_{\mu\nu\sigma\tau}C^{\mu\nu\sigma\tau} \m\alpha c_1^2 \m 2\,\alpha c_1 R + 3\,\alpha c_1 \dfrac{1}{\varphi^2} \partial_\mu \varphi \,\partial^\mu \varphi \Big)\nonumber\\
=&\int d^4x \sqrt{-g} \,\Big(\frac{1}{2\kappa^2}(R - 2\Lambda) + \beta \,C_{\mu\nu\sigma\tau}C^{\mu\nu\sigma\tau} \m \dfrac{1}{2} \partial_\mu \psi \,\partial^\mu \psi \Big)
\label{Sd}
\end{align} 
where $\Lambda=-\frac{c_1}{4}$ is the cosmological constant and $\frac{1}{2 \kappa^2}=-2\alpha\,c_1$ with $\kappa^2=8\,\pi\,G$ where $G$ is Newton's constant. We define $\psi =\frac{\sqrt{6}}{2\kappa}\,\ln \varphi$ so that the kinetic term for $\varphi$ is in canonical form.  Newton's constant and the cosmological constant can be chosen freely by adjusting the parameters $\alpha$ and $c_1$ (with $\alpha \,c_1 <0$ to ensure the correct sign for Newton's constant)\footnote{The constant $c_1$ is dimensionful and has units of (length)$^{-2}$. This stems from the fact that $c_1 \,\varphi$ in \reff{Sb} has units of (length)$^{-2}$ and $\varphi$ is assumed dimensionless in \reff{Conf}}. In AdS space, $\Lambda<0$, so that $c_1$ is positive and $\alpha$ is negative whereas in dS space, $\Lambda>0$ and $c_1$ is negative and $\alpha$ is positive. 

The conformal (Weyl) transformation \reff{Conf} is not valid for $\varphi=0$ and therefore the equivalence of the theories tacitly assumes a vacuum with $\varphi \neq 0$. This vacuum is not invariant under $\varphi \to \frac{\varphi}{\Omega^2}$ so that the restricted Weyl symmetry is spontaneously broken \cite{YNAE2}. This is evident from the fact that the final action \reff{Sd} now has an Einstein-Hilbert term with a cosmological constant. The massless scalar $\psi$ (defined above in terms of $\varphi$), is identified as the Nambu-Goldstone boson associated with the broken symmetry.      

When $\beta=0$, the resultant theory is a standard Einstein gravity (with a cosmological constant and a massless scalar field). We may couple it to the other matter fields such as Higgs field, gauge fields, and fermions, and one may even construct the standard model of particle physics coupled with gravity in our restricted Weyl invariant formulation \cite{YNAE2,YNAE1}. One of the goals of this paper is to study the effect of non-zero $\beta$.

\section{Critical gravity in the Einstein frame}

We now obtain the equations of motion and linearize them. We first rewrite the action \reff{Sd} in the form
\begin{align}
S_5=\frac{1}{2\kappa^2}\int d^4x \sqrt{-g} \,\Big(R - 2\Lambda + \gamma \,C_{\mu\nu\sigma\tau}C^{\mu\nu\sigma\tau} \m \kappa^2 \partial_\mu \psi \,\partial^\mu \psi \Big)
\label{Se}
\end{align}
where $\gamma=2\beta \kappa^2$ is now a dimensionful constant. The equations of motion are
\begin{align}
R_{\mu\nu} -\frac{1}{2} g_{\mu\nu} R + \Lambda g_{\mu\nu} + \gamma \Big(&-\frac{4}{3} R R_{\mu\nu} +\frac{1}{3}g_{\mu\nu}R^2 -\frac{1}{3}g_{\mu\nu}\Box R \nonumber \\&-\frac{2}{3}\nabla_{\mu}\nabla_{\nu} R +2 \Box R_{\mu\nu} +4 R_{\mu\sigma\nu\rho}R^{\sigma \rho}-g_{\mu\nu}R_{\sigma \rho}R^{\sigma\rho}\Big)\nonumber\\&+\kappa^2 \Big(\frac{1}{2}g_{\mu\nu} \partial_{\alpha} \psi\partial^{\alpha}\psi -\partial_{\mu}\psi\partial_{\nu}\psi\Big)=0\nonumber\\
\Box \psi=0
\label{EOM}
\end{align} 
One can easily see that $\psi = \bar{\psi} = \mathrm{const}$ is a solution of the equations of motion and then dS or AdS space-time are solutions for the metric equation of motion depending on the sign of $\Lambda$.

We linearize about a  dS or AdS background (denoted by a bar) so that
\begin{align}
g_{\mu\nu} =\bar{g}_{\mu\nu} + h_{\mu\nu}\quad;\quad \psi=  \bar{\psi} +\delta \psi
\label{delta}
\end{align} 
The Ricci scalar, Ricci tensor and Riemann tensor in a dS or AdS background are given by the following relations   
\beq
\bar{R}=4 \Lambda \quad;\quad \bar{R}_{\mu\nu} = \Lambda \bar{g}_{\mu\nu}\quad; \quad\bar{R}_{\rho\sigma\mu\nu}=\dfrac{\Lambda}{3}(\bar{g}_{\rho\mu}\bar{g}_{\sigma\nu}-\bar{g}_{\rho\nu}\bar{g}_{\sigma\mu})\,.
\eeq{Rbar}
We work in harmonic gauge
\beq
\bar{\nabla}_{\nu}h= \bar{\nabla}_{\alpha} h^{\alpha}_{\nu}
\eeq{harmonic}
where $h = h_{\mu\nu} \bar{g}^{\mu\nu} $ is the trace of $h_{\mu\nu}$. The linearized equations to first order in $h_{\mu\nu}$ and $\delta \psi$ are (see Appendix A)
\begin{align}
&\dfrac{1}{2}\bar{\nabla}_{\mu}\bar{\nabla}_{\nu} h +\dfrac{1}{3}\Lambda h_{\mu\nu} +\dfrac{1}{6}\Lambda\bar{g}_{\mu\nu}h -\dfrac{1}{2} \bar{\Box}h_{\mu\nu} \nonumber\\ +&\gamma\Big(-\dfrac{2}{3}\Lambda \bar{\nabla}_{\mu}\bar{\nabla}_{\nu} h -\dfrac{8}{9} \Lambda^2 h_{\mu\nu}+\dfrac{2}{9}\Lambda^2\bar{g}_{\mu\nu}h +2\Lambda\bar{\Box}h_{\mu\nu} 
\nonumber\\&-\dfrac{1}{3}\Lambda \bar{g}_{\mu\nu} \bar{\Box}h + \bar{\Box}\bar{\nabla}_{\mu}\bar{\nabla}_{\nu} h -\bar{\Box}^2 h_{\mu\nu}\Big)= 0\nonumber\\
\bar{\Box} \delta \psi=0\,.
\label{linearize}
\end{align}
Contracting the above equation yields $\Lambda\,h=0$. Since 
$\Lambda \neq 0$, we obtain $h=0$. For $h=0$, the harmonic gauge condition \reff{harmonic} yields $\bar{\nabla}_{\alpha} h^{\alpha}_{\nu}=0$. Therefore $h_{\mu\nu}$ is both traceless and transverse. With $h=0$, the linearized equations reduce to
\begin{align}
\dfrac{1}{3}\Lambda h_{\mu\nu} -\dfrac{1}{2} \bar{\Box}h_{\mu\nu} +\gamma\Big(-\dfrac{8}{9} \Lambda^2 h_{\mu\nu} +2\Lambda\bar{\Box}h_{\mu\nu} -\bar{\Box}^2 h_{\mu\nu}\Big)= 0\nonumber\\
\bar{\Box} \delta \psi=0\,.
\label{linearizeA}
\end{align}
We can rewrite the above in the following form
\begin{align}
-\gamma\Big(\bar{\Box}-\dfrac{2\Lambda}{3}\Big)\Big(\bar{\Box}-\dfrac{4\Lambda}{3} +\dfrac{1}{2\gamma}\Big)h_{\mu\nu}=0\nonumber\\
\bar{\Box} \delta\psi=0\,.
\label{linearizeB}
\end{align}
For generic parameters, the above equations describe a massless graviton (which obey $(\bar{\Box}-\frac{2\Lambda}{3}) h_{\mu\nu}=0$) and a massless Nambu-Goldstone scalar $ \delta\psi$. They also describe a ``massive" spin two excitation that satisfies $\Big(\bar{\Box}-\frac{4\Lambda}{3} +\frac{1}{2\gamma}\Big)\,h_{\mu\nu} = 0$. In particular, it has negative mass squared when $-\frac{4\Lambda}{3} +\frac{1}{2\gamma}>-\frac{2\Lambda}{3}$, but is stable in the sense that the time dependence of the mode $\sim e^{-i\Delta t}$ with $\Delta = \frac{3\pm\sqrt{9-12\frac{m^2}{\Lambda}}}{2}$ is not exponentially growing when $m^2 = \frac{2\Lambda}{3} - \frac{1}{2\gamma} \ge \frac{3\Lambda}{4}$ (i.e. $\gamma \le 0$ or $\gamma \ge -\frac{6}{\Lambda}$) in the global AdS space-time \cite{PopeLu2}.\footnote{While the time dependence of this excitation is not exponentially growing in the global AdS space, the representation is not unitary: if we assume the lowest energy spin two mode has a positive norm, the descendant mode has a negative norm from the representation theory of the AdS algebra, so we cannot quantize the excitation while keeping the unitarity. Since we have a worse problem, i.e. the negative energy or negative norm for the lowest energy excitation, this just adds a small complication. The only resolution that we can propose here is that we discard these excitations as we will do in the following.}

At the critical value $\frac{1}{2\gamma}=\frac{2\Lambda}{3}$ (or $\gamma=\frac{3}{4\Lambda}$), the ``massive" spin two excitations become massless and we also have degenerate massless gravitons accompanied by logarithmic modes. The resulting theory is known as critical gravity in four dimensions. Substituting $\gamma =2\beta \kappa^2$, $\kappa^2=-\frac{1}{4 \alpha c_1}$ and $\Lambda=-\frac{c_1}{4}$  the critical condition becomes $\beta=6 \alpha$ where $\alpha$ and $\beta$ are the parameters in the original action \reff{Rw1}. In the next section, we obtain the critical condition $\beta=6\alpha$ directly from the original higher-derivative action \reff{Rw1}.

About the AdS or dS vacua, either one of massive spin two excitations or massless spin two excitations must have negative energy and hence they are ghosts. The behavior i.e. which one becomes a ghost changes at the critical value of $\gamma =\frac{3}{4\Lambda}$ or $\beta=6\alpha$. When $\gamma \ge \frac{3}{4\Lambda}$ or $\beta \ge 6\alpha$, the massless graviton has positive energy. In Euclidean AdS or in dS space-time, it was pointed out by Maldacena \cite{Maldacena} (see also \cite{PopeLu2}) that one may impose boundary conditions to remove the ghost degrees of freedom. For physical applications, we would like the massless spin two excitations to have positive energy.\footnote{In the other regime $\beta < 6\alpha$, massless gravitons have negative energy, so it is excluded here.}

In pseudo-Rieamannian space-time, a truncation of the spectrum by a boundary condition requires a careful dynamical consideration.
While in \cite{PopeLu2} it seems that they have the Lorentzian AdS space-time in mind, in the original paper by Maldacena \cite{Maldacena} he restricted himself to the situations in Euclidean AdS or dS space-time.  This was because in the Lorentzian AdS space-time the mode that we would like to truncate becomes normalizable and could appear under the interaction. In this paper, we focus on the conservative situations Maldacena considered.

Near the (Euclidean) AdS boundary (at $z=0$), the metric fluctuation behaves as $\sim \frac{\delta h_{\mu\nu}}{z^\Delta}$, where $\Delta = \frac{3\pm\sqrt{9-12\frac{m^2}{\Lambda}}}{2}$, so compared to the massless graviton with $m^2=0$, the ``massive" mode with $ \frac{3\Lambda }{4} \le m^2 <0$ decays more slowly near $z=0$ and can be truncated by imposing the boundary conditions. For our case, this is possible when $\gamma \ge -\frac{6}{\Lambda}$ or $\beta \ge -48\alpha$ (including the limiting case of conformal gravity at $\gamma = \infty $ or $\alpha = 0$ with the partial massless spectrum \cite{Deser:2001pe}), at which the ghost mode decays more slowly at the Euclidean AdS boundary.

At the critical value of $\beta = 6\alpha$, one may still remove the logarithmic excitations, but it turns out that the resulting gravitational theory is rather empty except for the Nambu-Goldstone mode. Without imposing the boundary conditions, it contains ghosts but it may have certain applications to holographic three-dimensional logarithmic conformal field theories \cite{Hogervorst:2016itc}.  

Similarly near the dS future boundary (at $\eta=0$), the metric fluctuation behaves as $\sim \frac{\delta{h}_{\mu\nu}}{\eta^{\Delta}}$ with  $\Delta = \frac{3\pm\sqrt{9-12\frac{m^2}{\Lambda}}}{2}$. Compared to the massless graviton with $m^2=0$, the ``massive" spin two mode decays more slowly when $m^2 \ge 0$ (i.e. $\beta < 6\alpha$), in which case we may put the (future) boundary conditions to remove the ghost massive spin two mode.

\section{Critical gravity from the original higher-derivative action}

The equations of motion for the original action \reff{Rw1} are
\begin{align}
(2 \alpha-\dfrac{4\beta}{3}) RR_{\mu\nu} +(\dfrac{\beta}{3}-\dfrac{\alpha}{2})g_{\mu\nu}R^2&+(2\alpha-\dfrac{\beta}{3})g_{\mu\nu}\Box R -(2 \alpha +\dfrac{2\beta}{3})\nabla_{\mu}\nabla_{\nu} R + 2 \beta \Box R_{\mu\nu} \nonumber \\& +4 \beta R_{\mu\sigma\nu\rho}R^{\sigma \rho}-\beta g_{\mu\nu} R_{\sigma \rho}R^{\sigma\rho}=0\,.
\label{EOM3}
\end{align}
We consider a dS or AdS background where the Ricci scalar, Ricci tensor and Riemann tensor are given by \reff{Rbar}. The linearized equations in harmonic gauge \reff{harmonic} are (see Appendix A)
\begin{align}
&-\beta \bar{\Box}^2 h_{\mu\nu} +(2\beta-4\alpha)\Lambda\bar{\Box}h_{\mu\nu}+(\dfrac{8\alpha}{3}-\dfrac{8 \beta}{9}) \Lambda^2 h_{\mu\nu} + (6\alpha-\dfrac{2\beta}{3})\Lambda \bar{\nabla}_{\mu}\bar{\nabla}_{\nu} h \nonumber\\&+(\dfrac{2\beta}{9}-\dfrac{2\alpha}{3})\Lambda^2\bar{g}_{\mu\nu}h -(2\alpha+\dfrac{\beta}{3})\Lambda \bar{g}_{\mu\nu} \bar{\Box}h+ \beta \bar{\Box}\bar{\nabla}_{\mu}\bar{\nabla}_{\nu} h =0\,.
\label{linearize2}
\end{align} 
Contraction yields $-6 \,\alpha \Lambda \bar{\Box}h=0$. If $\alpha\neq 0$ (i.e. non-conformal gravity) and $\Lambda \neq 0$ then we obtain 
\beq
\bar{\Box} h=0 \,.
\eeq{Boxh} 
Consequently, non-conformal scale invariant gravity in an dS or AdS background has a propagating massless scalar field. We assume $\alpha \neq 0$ and substitute $\bar{\Box} h=0$ in \reff{linearize2}. It is convenient to define 
\beq
\tilde{h}_{\mu\nu}=h_{\mu\nu} -\dfrac{1}{4}\bar{g}_{\mu\nu}h -\dfrac{3}{4\Lambda}\bar{\nabla}_{\mu}\bar{\nabla}_{\nu} h \,. 
\eeq{htilde}
Note that $\tilde{h}_{\mu\nu}$ is traceless and transverse. It is traceless because in contracting the above $\bar{\Box} h=0$. It is transverse because
\begin{align}
\bar{\nabla}^{\mu}\tilde{h}_{\mu\nu}&=\bar{\nabla}^{\mu}h_{\mu\nu} -\dfrac{1}{4}\bar{\nabla}_{\nu}h -\dfrac{3}{4\Lambda}\bar{\Box}\bar{\nabla}_{\nu} h\nonumber\\& = \bar{\nabla}_{\nu}h
-\dfrac{1}{4}\bar{\nabla}_{\nu}h -\dfrac{3}{4\Lambda}(\Lambda \bar{\nabla}_{\nu} h)=0
\label{transverse2}
\end{align}
where we used the harmonic gauge condition \reff{harmonic} together with the relation   
\beq
\bar{\Box}\bar{\nabla}_{\nu} h = \bar{\nabla}_{\nu} \bar{\Box} h + \Lambda \bar{\nabla}_{\nu} h=\Lambda \bar{\nabla}_{\nu} h\,.
\eeq{covariant1}
The above was obtained using the commutation relations between covariant derivatives \reff{Covariant} together with the expression \reff{Rbar} for the Riemann tensor in a dS or AdS background. The result \reff{Boxh} was also applied. The linearized equations \reff{linearize2} expressed in terms of the new field \reff{htilde} reduce to
\begin{align}
&-\beta \bar{\Box}^2 \tilde{h}_{\mu\nu} +(2\beta-4\alpha)\Lambda\bar{\Box}\tilde{h}_{\mu\nu}+(\dfrac{8\alpha}{3}-\dfrac{8 \beta}{9}) \Lambda^2 \tilde{h}_{\mu\nu} =0
\label{linearize3}
\end{align} 
where we used
\beq 
\bar{\Box}\bar{\nabla}_{\mu}\bar{\nabla}_{\nu} h=\bar{\nabla}_{\mu}\bar{\nabla}_{\nu}\bar{\Box} h -\dfrac{2\Lambda}{3}\bar{g}_{\mu\nu} \bar{\Box} h +\dfrac{8\Lambda}{3}\bar{\nabla}_{\mu}\bar{\nabla}_{\nu} h=\dfrac{8\Lambda}{3}\bar{\nabla}_{\mu}\bar{\nabla}_{\nu} h\,.
\eeq{BoxNabla2}
The above was again obtained using the commutation relations between covariant derivatives \reff{Covariant} together with the expression \reff{Rbar} for the Riemann tensor in a dS or AdS background. The result \reff{Boxh} was also applied.
The linearized equations \reff{linearize3} can be written in the following form
\beq
-\beta(\bar{\Box}-\dfrac{2\Lambda}{3})(\bar{\Box}-\dfrac{4\Lambda}{3} +\dfrac{4 \alpha}{\beta} \Lambda)\tilde{h}_{\mu\nu}= 0\,.
\eeq{critical}
Then  $(\bar{\Box}-\dfrac{4\Lambda}{3} +\dfrac{4 \alpha}{\beta} \Lambda) \to (\bar{\Box}-\dfrac{2\Lambda}{3})$ at the critical value of $\beta = 6 \alpha$. This agrees with the critical value obtained in the previous section.  At the critical value, the linearized equations become
\beq
(\bar{\Box}-\dfrac{2\Lambda}{3})(\bar{\Box}-\dfrac{2\Lambda}{3})\tilde{h}_{\mu\nu}= 0\,.
\eeq{critical2}

The necessity and the origin of the field redefinition from $h_{\mu\nu}$ to $\tilde{h}_{\mu\nu}$ is as follows. We assumed the same harmonic gauge in both formulations. The two formulations are related by the conformal transformation \reff{Conf}, but the conformal transformation does not preserve the harmonic gauge condition, so we cannot regard $h_{\mu\nu}$ as the same excitations. Rather they must be identified after the further gauge transformation $\delta h_{\mu\nu} = \nabla_\mu v_\nu + \nabla_\nu v_\mu$. Here, the necessary diffeomorphism is given by $v_\mu \propto \nabla_\mu h$.

\section{Energy and entropy of Schwarzschild and Kerr AdS or dS black hole and the action at the critical condition}
The energy of a Schwarzschild or Kerr AdS or dS black hole for the same action \reff{Rw1} can be found in \cite{Adami}. In our notation it is given by  
\beq
E=\Lambda\Big(8 \alpha - \dfrac{4}{3} \beta\Big)\int d^3x \sqrt{-\bar{g}} \bar{\xi}_{\nu} \mathcal{G}^{0\nu}_{(1)}
\eeq{energy} 
where $\mathcal{G}^{0\nu}_{(1)}$ represents the linearization of components of the Einstein tensor, $\bar{\xi}_{\nu}$ represents Killing symmetries with the bar denoting the AdS or dS background (see \cite{Adami} for details). Regardless of the value of the integral, the energy vanishes at the critical condition $\beta=6\alpha$. 

The entropy of the Schwarzschild or Kerr AdS or dS black hole can be evaluated using the Wald entropy formula \cite{Wald,Jacobson}
\beq
S=-2\pi \oint_{\Sigma}\Big(\dfrac{\delta \mathcal{L}}{\delta R_{abcd}}\Big)^{(0)} \hat{\epsilon}_{ab} \hat{\epsilon}_{cd} \bar{\epsilon}
\eeq{Wald}
where the integral is on the 2-dimensional spacelike bifurcation surface $\Sigma$, $\hat{\epsilon}_{ab}$ is the binormal vector to the bifurcation surface (it is normalized such that $\hat{\epsilon}_{ab}\hat{\epsilon}^{ab}=-2$). The quantity $\bar{\epsilon}$ is the induced volume form on the bifurcation surface. The $(0)$ superscript in $(\tfrac{\delta \mathcal{L}}{\delta R_{abcd}})^{(0)}$ means that the partial derivative is evaluated on the solution of the equations of motion. For both the Schwarzschild or Kerr AdS or dS black hole this means evaluating the derivative at $R=4 \Lambda$ and $R_{\mu\nu}= \Lambda g_{\mu\nu}$. For the quadratic action \reff{Rw1} a straightforward calculation yields
\beq
S=-2\pi\Lambda\Big(8 \alpha - \dfrac{4}{3} \beta\Big)\oint_{\Sigma}\,g^{ac}g^{bd}\hat{\epsilon}_{ab} \hat{\epsilon}_{cd} \bar{\epsilon}\,.
\eeq{Wald2}
Again, regardless of the value of the integral, the entropy vanishes at the critical condition $\beta=6\alpha$.  

So we see that both the entropy and energy of a Schwarzschild or Kerr AdS or dS black hole vanish at the critical condition in agreement with the results of \cite{PopeLu1}.They agree even though in our case the energy and entropy formulas above are for purely higher-derivative gravity and do not include a contribution from Einstein gravity in contrast to the case in \cite{PopeLu1}.

It had been pointed out in \cite{Deser} that the energy vanishes at $\beta =6 \alpha$ (in their notation this was $\beta=-4\alpha$) and that this leads to an action which is proportional to the square of the trace free part of the Ricci tensor i.e. the square of $\tilde{R}_{\mu\nu}=R_{\mu\nu} -\frac{1}{4} g_{\mu\nu} R$. So it is worth noting that our critical condition obtained in the context of critical gravity yields a final action which is proportional to the square of $\tilde{R}_{\mu\nu}$.  

\section{Conclusion}

In the original study of critical gravity in four dimensions, Einstein gravity with a cosmological constant were added to higher-derivative gravity. In this work we showed that critical gravity can occur in four dimensional scale invariant gravity, a purely quadratic action where there is no Einstein-Hilbert term or a cosmological constant. We showed in two independent ways that the critical condition is $\beta=6 \alpha$ where $\alpha$ and $\beta$ are dimensionless parameters appearing in action \reff{Rw1}. The formulas for the energy and entropy of a Schwarzschild or Kerr AdS or dS black hole stemming from a purely quadratic action yields zero at the critical condition, the same value obtained in \cite{PopeLu1} (even though zero is obtained from different contributions in the two cases). At the critical condition, the action becomes proportional to the square of the trace free part of the Ricci tensor something that had been noticed in a separate context by Deser and Tekin in their study of energy in actions containing quadratic gravity \cite{Deser}.  

We elucidated the role that boundary conditions can play \cite{Maldacena, PopeLu2}. Boundary conditions are important for two reasons. In critical gravity there remains logarithmic spin two modes that are ghosts. Secondly, in critical gravity the massless gravitons yield zero energy so the theory is sort of empty. We found that in (Euclidean) AdS space, one can impose boundary conditions and obtain a unitary solution with positive energy for the case $\beta \ge -48 \alpha$. In dS space, after imposing boundary conditions, positive energy solutions could be obtained for the case $\beta<6 \alpha$. 

There are a couple of future directions to be pursued. Our scale invariant gravity is power-counting renormalizable without the boundary conditions. It is an interesting future question to see if it is still renormalizable with the specific boundary conditions. The background solution and the boundary condition depend on the parameters of the theory, so it is a non-trivial question to see how the renormalization group running of parameters are compatible with each other. In this work, the massless scalar degrees of freedom, which is associated with the Nambu-Goldstone mode for the spontaneous symmetry breaking, does not play an important role, but we may expect there are non-trivial solutions sourced by it. 

\section*{Acknowledgements}
The work by Y.N  is in part supported by JSPS KAKENHI Grant Number 17K14301. A.E. is supported by a discovery grant of the Natural Sciences and Engineering Research Council of Canada (NSERC).

\begin{appendices}
\numberwithin{equation}{section}
\setcounter{equation}{0}
\section{Linearized equations about a dS or AdS background}
We linearize the metric about a curved background (denoted by a bar) so that
\beq
g_{\mu\nu} =\bar{g}_{\mu\nu} + h_{\mu\nu} \,.
\eeq{metric2}
where $h_{\mu\nu}$ is the perturbation. Linearized quantities or deviations from the background are denoted with a $\delta$ e.g. $R_{\mu\nu} = \bar{R}_{\mu\nu} + \delta R_{\mu\nu}\,;\,R = \bar{R} + \delta R$, etc.
We have the following useful results for any general background
\beq
\delta \Gamma^{\lambda}_{\beta\mu}=\dfrac{1}{2}\bar{g}^{\lambda \tau}(\bar{\nabla}_{\beta}h_{\mu\tau}+\bar{\nabla}_{\mu}h_{\beta\tau}-\bar{\nabla}_{\tau}h_{\beta\mu})
\eeq{Christoffel}
\begin{align}
\delta R^{\alpha}_{\,\beta \mu\nu}&=\bar{\nabla}_{\mu}\delta \Gamma^{\alpha}_{\beta\nu}-\bar{\nabla}_{\nu}\delta \Gamma^{\alpha}_{\beta\mu}\nonumber\\
\delta R_{\mu\nu} &= \dfrac{1}{2}(\bar{\nabla}^{\alpha}\bar{\nabla}_{\mu} h_{\alpha\nu} +\bar{\nabla}^{\alpha}\bar{\nabla}_{\nu} h_{\alpha\mu} -\bar{\nabla}_{\mu}\bar{\nabla}_{\nu} h -\bar{\Box} h_{\mu\nu})\nonumber\\
\delta R&= -\bar{R}^{\mu\nu} h_{\mu\nu} + 
\bar{\nabla}^{\mu}\bar{\nabla}^{\nu} h_{\mu\nu} -\bar{\Box} h \,.
\label{deltaR}
\end{align}
For a dS or AdS background, the Ricci scalar, Ricci tensor and Riemann tensor are given by \reff{Rbar}. We work in harmonic gauge \reff{harmonic}. We make use of the following commutation relations between covariant derivatives
\beq
[\nabla_{\sigma},\nabla_{\rho}] T_{abc....}= -\bar{R}_{\sigma\rho \ a}^{\quad j} T_{jbc...} -\bar{R}_{\sigma\rho \ b }^{\quad j} T_{ajc...} -\bar{R}_{\sigma\rho \ c}^{\quad j} T_{abj...} +... 
\eeq{Covariant} 
Using the harmonic gauge and the above commutation relations we obtain in a dS or AdS background the following results 
\begin{align}
\delta R&=- \Lambda h\nonumber\\ 
\delta R_{\mu\nu}&=\dfrac{1}{2}(\bar{\nabla}_{\mu}\bar{\nabla}_{\nu} h +\dfrac{8 \Lambda}{3} h_{\mu\nu}-\dfrac{2\Lambda}{3}\bar{g}_{\mu\nu}h -\bar{\Box} h_{\mu\nu})\nonumber\\
\delta R^{\alpha}_{\,\beta \mu\nu}&=
\dfrac{1}{2}\Big(\bar{\nabla}_{\mu}\bar{\nabla}_{\beta} h^{\alpha}_{\nu} -\bar{\nabla}_{\mu}\bar{\nabla}^{\alpha} h_{\beta\nu}-\bar{\nabla}_{\nu}\bar{\nabla}^{\beta} h^{\alpha}_{\mu} +\bar{\nabla}_{\nu}\bar{\nabla}^{\alpha} h_{\beta\mu}\Big) \nonumber\\&+\dfrac{\Lambda}{6}
\Big(-\bar{g}_{\beta\nu} h^{\alpha}_{\mu} +\bar{g}_{\beta\mu} h^{\alpha}_{\nu} -\bar{g}^{\alpha}_{\nu} h_{\mu\beta} +\bar{g}^{\alpha}_{\mu} h_{\nu\beta}\Big)\,.
\label{DeltaRiemann}
\end{align}
Using \reff{DeltaRiemann} and \reff{Christoffel} we gather below some useful results in harmonic gauge that enter the linearized equations: 
\begin{align}
\bar{g}^{\alpha\beta}\bar{\nabla}_{\alpha}
(-\delta \Gamma^{\lambda}_{\beta\mu} \bar{R}_{\lambda \nu}-\delta \Gamma^{\lambda}_{\beta\nu} \bar{R}_{\lambda \mu})&= -\Lambda \bar{\Box} h_{\mu\nu}\nonumber\\
\delta R_{\mu\sigma\nu\rho} \bar{R}^{\sigma \rho}&= \Lambda^2 h_{\mu\nu} +\dfrac{1}{2}\Lambda\bar{\nabla}_{\mu}\bar{\nabla}_{\nu} h-\dfrac{1}{2} \Lambda \bar{\Box} h_{\mu\nu}\nonumber\\
\bar{R}_{\mu\sigma\nu\rho} \,\delta R^{\sigma \rho} &=-\dfrac{5}{9}\Lambda^2\bar{g}_{\mu\nu}h+\dfrac{2}{9}\Lambda^2 h_{\mu\nu}-\dfrac{1}{6}\Lambda\bar{\nabla}_{\mu}\bar{\nabla}_{\nu} h +\dfrac{1}{6} \Lambda \bar{\Box} h_{\mu\nu}\nonumber \\
\bar{g}_{\mu\nu}(\delta R^{\sigma \rho}\bar{R}_{\sigma \rho} +
\bar{R}^{\sigma \rho}\,\delta R_{\sigma \rho})&=-2\Lambda^2\bar{g}_{\mu\nu}h\,.
\label{Gather}
\end{align}
We will also make use of
\beq
\delta(\Box R_{\mu\nu})= \bar{\Box}\delta R_{\mu\nu} + \bar{g}^{\alpha\beta}\bar{\nabla}_{\alpha}
(-\delta \Gamma^{\lambda}_{\beta\mu} \bar{R}_{\lambda \nu}-\delta \Gamma^{\lambda}_{\beta\nu} \bar{R}_{\lambda \mu})\,.
\eeq{DeltaBoxR}
The linearized equations corresponding to the equations of motion \reff{EOM} in section 3 to first order in $h_{\mu\nu}$ and $\delta \psi$ are
\begin{align}
&\delta R_{\mu\nu} -\frac{1}{2} h_{\mu\nu} \bar{R}-\frac{1}{2} \bar{g}_{\mu\nu} \delta R  + \Lambda h_{\mu\nu} + \gamma \Big(-\frac{4}{3} (\delta R \bar{R}_{\mu\nu} + \bar{R} \delta R_{\mu\nu}) +\frac{1}{3}h_{\mu\nu}\bar{R}^2 \nonumber\\&+\dfrac{2}{3} \bar{g}_{\mu\nu}\bar{R}\delta R -\frac{1}{3}\bar{g}_{\mu\nu}\bar{\Box}\delta R -\frac{2}{3}\bar{\nabla}_{\mu}\bar{\nabla}_{\nu} \delta R +2 \bar{\Box} \delta R_{\mu\nu} + 2\bar{g}^{\alpha\beta}\bar{\nabla}_{\alpha}
(-\delta \Gamma^{\lambda}_{\beta\mu} \bar{R}_{\lambda \nu}-\delta \Gamma^{\lambda}_{\beta\nu} \bar{R}_{\lambda \mu})\nonumber\\& +4 \delta R_{\mu\sigma\nu\rho}\bar{R}^{\sigma \rho}+4 \bar{R}_{\mu\sigma\nu\rho}\delta R^{\sigma \rho}-h_{\mu\nu}\bar{R}_{\sigma \rho}\bar{R}^{\sigma\rho}-\bar{g}_{\mu\nu}(\bar{R}_{\sigma \rho}\delta R^{\sigma\rho} +\delta R_{\sigma \rho}\bar{R}^{\sigma\rho})\Big)=0\nonumber\\\nonumber\\
&\bar{\Box} \delta \psi=0\,
\label{EOM2}
\end{align} 
where we made use of \reff{DeltaBoxR}. Using the results \reff{DeltaRiemann} and \reff{Gather}, the linearized equations become
\begin{align}
&\dfrac{1}{2}\bar{\nabla}_{\mu}\bar{\nabla}_{\nu} h +\dfrac{1}{3}\Lambda h_{\mu\nu} +\dfrac{1}{6}\Lambda\bar{g}_{\mu\nu}h -\dfrac{1}{2} \bar{\Box}h_{\mu\nu} \nonumber\\ +&\gamma\Big(-\dfrac{2}{3}\Lambda \bar{\nabla}_{\mu}\bar{\nabla}_{\nu} h -\dfrac{8}{9} \Lambda^2 h_{\mu\nu}+\dfrac{2}{9}\Lambda^2\bar{g}_{\mu\nu}h +2\Lambda\bar{\Box}h_{\mu\nu} 
\nonumber\\&-\dfrac{1}{3}\Lambda \bar{g}_{\mu\nu} \bar{\Box}h + \bar{\Box}\bar{\nabla}_{\mu}\bar{\nabla}_{\nu} h -\bar{\Box}^2 h_{\mu\nu}\Big)= 0\nonumber\\
\bar{\Box} \delta \psi=0
\label{linearize4}
\end{align} 
which are the equations quoted in \reff{linearize} in section 3.

The linearized equations corresponding to the equations of motion \reff{EOM3} in section 4 to first order in $h_{\mu\nu}$ are
\begin{align}
&(2 \alpha-\dfrac{4\beta}{3}) (\bar{R}\delta R_{\mu\nu}+\delta R \bar{R}_{\mu\nu})+(\dfrac{\beta}{3}-\dfrac{\alpha}{2})(h_{\mu\nu}\bar{R}^2+2 \bar{g}_{\mu\nu}\bar{R} \delta R)+(2\alpha-\dfrac{\beta}{3})\bar{g}_{\mu\nu} \bar{\Box}\delta R \nonumber\\&-(2 \alpha +\dfrac{2\beta}{3})\bar{\nabla}_{\mu}\bar{\nabla}_{\nu} \delta R + 2 \beta (\bar{\Box} \delta R_{\mu\nu} + \bar{g}^{\alpha\beta}\bar{\nabla}_{\alpha}
(-\delta \Gamma^{\lambda}_{\beta\mu} \bar{R}_{\lambda \nu}-\delta \Gamma^{\lambda}_{\beta\nu} \bar{R}_{\lambda \mu}))\nonumber \\& +4 \beta (\delta R_{\mu\sigma\nu\rho}\bar{R}^{\sigma \rho}+\bar{R}_{\mu\sigma\nu\rho}\delta R^{\sigma \rho})-\beta h_{\mu\nu} \bar{R}_{\sigma \rho}\bar{R}^{\sigma\rho}-\beta \bar{g}_{\mu\nu}( \bar{R}_{\sigma \rho}\delta R^{\sigma\rho}+\delta R_{\sigma \rho}\bar{R}^{\sigma\rho})=0\,.
\label{EOM4}
\end{align}
Again, using the results \reff{DeltaRiemann} and \reff{Gather}, the linearized equations become
\begin{align}
&-\beta \bar{\Box}^2 h_{\mu\nu} +(2\beta-4\alpha)\Lambda\bar{\Box}h_{\mu\nu}+(\dfrac{8\alpha}{3}-\dfrac{8 \beta}{9}) \Lambda^2 h_{\mu\nu} + (6\alpha-\dfrac{2\beta}{3})\Lambda \bar{\nabla}_{\mu}\bar{\nabla}_{\nu} h \nonumber\\&+(\dfrac{2\beta}{9}-\dfrac{2\alpha}{3})\Lambda^2\bar{g}_{\mu\nu}h -(2\alpha+\dfrac{\beta}{3})\Lambda \bar{g}_{\mu\nu} \bar{\Box}h+ \beta \bar{\Box}\bar{\nabla}_{\mu}\bar{\nabla}_{\nu} h =0\,.
\label{linearize6}
\end{align}
which are the equations quoted in \reff{linearize2} in section 4.
\end{appendices}

\end{document}